\documentclass[
reprint,
superscriptaddress,
 amsmath,amssymb,
aps,
pra,
]{revtex4-2}

\usepackage{times} 
\usepackage{graphicx}
\usepackage{dcolumn}
\usepackage{bm}
\usepackage{siunitx}
\usepackage{hyperref}
\usepackage{adjustbox}
\usepackage{comment}
\usepackage{tabularx}

\begin{document}

\preprint{PRL/123-QED}

\title{2D transverse laser cooling of a hexapole focused beam of cold BaF molecules}

\author{J.W.F.~van Hofslot}
\affiliation{Van Swinderen Institute for Particle Physics and Gravity, University of Groningen, The Netherlands}%
\affiliation{Nikhef, National Institute for Subatomic Physics, Amsterdam, The Netherlands}

\author{I.E.~Thompson}%
\affiliation{Van Swinderen Institute for Particle Physics and Gravity, University of Groningen, The Netherlands}%

\author{A.~Touwen}
\affiliation{Van Swinderen Institute for Particle Physics and Gravity, University of Groningen, The Netherlands}%
\affiliation{Nikhef, National Institute for Subatomic Physics, Amsterdam, The Netherlands}
\affiliation{Department of Physics and Astronomy, and LaserLaB, Vrije Universiteit Amsterdam, The Netherlands}

\author{N.~Balasubramanian}
\affiliation{Van Swinderen Institute for Particle Physics and Gravity, University of Groningen, The Netherlands}%
\affiliation{Nikhef, National Institute for Subatomic Physics, Amsterdam, The Netherlands}

\author{R.~Bause}
\thanks{Present address: Menlo Systems GmbH, 82152 Martinsried, Germany}
\affiliation{Van Swinderen Institute for Particle Physics and Gravity, University of Groningen, The Netherlands}%
\affiliation{Nikhef, National Institute for Subatomic Physics, Amsterdam, The Netherlands}

\author{H.L.~Bethlem}%
\affiliation{Van Swinderen Institute for Particle Physics and Gravity, University of Groningen, The Netherlands}%
\affiliation{Department of Physics and Astronomy, and LaserLaB, Vrije Universiteit Amsterdam, The Netherlands}

\author{A.~Borschevsky}
\affiliation{Van Swinderen Institute for Particle Physics and Gravity, University of Groningen, The Netherlands}%
\affiliation{Nikhef, National Institute for Subatomic Physics, Amsterdam, The Netherlands}

\author{T.H.~Fikkers}
\affiliation{Van Swinderen Institute for Particle Physics and Gravity, University of Groningen, The Netherlands}%
\affiliation{Nikhef, National Institute for Subatomic Physics, Amsterdam, The Netherlands}

\author{S.~Hoekstra}
\thanks{Corresponding author: s.hoekstra@rug.nl} 
\affiliation{Van Swinderen Institute for Particle Physics and Gravity, University of Groningen, The Netherlands}%
\affiliation{Nikhef, National Institute for Subatomic Physics, Amsterdam, The Netherlands}

\author{S.A.~Jones}
\affiliation{Van Swinderen Institute for Particle Physics and Gravity, University of Groningen, The Netherlands}%
\affiliation{Nikhef, National Institute for Subatomic Physics, Amsterdam, The Netherlands}

\author{J.E.J.~Levenga}
\affiliation{Van Swinderen Institute for Particle Physics and Gravity, University of Groningen, The Netherlands}%
\affiliation{Nikhef, National Institute for Subatomic Physics, Amsterdam, The Netherlands}

\author{M.C.~Mooij}
\affiliation{Nikhef, National Institute for Subatomic Physics, Amsterdam, The Netherlands}
\affiliation{Department of Physics and Astronomy, and LaserLaB, Vrije Universiteit Amsterdam, The Netherlands}

\author{H.~Mulder}
\affiliation{Van Swinderen Institute for Particle Physics and Gravity, University of Groningen, The Netherlands}%
\affiliation{Nikhef, National Institute for Subatomic Physics, Amsterdam, The Netherlands}

\author{B.A.~Nijman}
\affiliation{Van Swinderen Institute for Particle Physics and Gravity, University of Groningen, The Netherlands}%
\affiliation{Nikhef, National Institute for Subatomic Physics, Amsterdam, The Netherlands}

\author{E.H.~Prinsen}
\affiliation{Van Swinderen Institute for Particle Physics and Gravity, University of Groningen, The Netherlands}%
\affiliation{Nikhef, National Institute for Subatomic Physics, Amsterdam, The Netherlands}

\author{B.J.~Schellenberg}
\affiliation{Van Swinderen Institute for Particle Physics and Gravity, University of Groningen, The Netherlands}%
\affiliation{Nikhef, National Institute for Subatomic Physics, Amsterdam, The Netherlands}

\author{L.~van Sloten}
\affiliation{Van Swinderen Institute for Particle Physics and Gravity, University of Groningen, The Netherlands}%
\affiliation{Nikhef, National Institute for Subatomic Physics, Amsterdam, The Netherlands}

\author{R.G.E.~Timmermans}
\affiliation{Van Swinderen Institute for Particle Physics and Gravity, University of Groningen, The Netherlands}%
\affiliation{Nikhef, National Institute for Subatomic Physics, Amsterdam, The Netherlands}

\author{W.~Ubachs}
\affiliation{Department of Physics and Astronomy, and LaserLaB, Vrije Universiteit Amsterdam, The Netherlands}

\author{J.~de~Vries}
\affiliation{Van Swinderen Institute for Particle Physics and Gravity, University of Groningen, The Netherlands}%

\affiliation{Institute of Physics and Delta Institute for Theoretical Physics, University of Amsterdam, The Netherlands}
\author{L.~Willmann,$^{1,2}$ for the NL-\textit{e}EDM collaboration}%


\date{\today}

\begin{abstract}
A cryogenic buffer gas beam, an electrostatic hexapole lens, and 2D transverse Doppler laser cooling are combined to produce a bright beam of barium monofluoride ($^{138}$Ba$^{19}$F) molecules. Experimental results and trajectory simulations are used to study the laser cooling effect as a function of laser detuning, laser power, laser alignment, and interaction time. A scattering rate of $6.1(1.4)\times10^{5}\ \unit{s^{-1}}$ on the laser cooling transition is obtained; this is $14 \%$ of the expected maximum, which is attributed to limited control of the magnetic field used to remix dark states. Using 3 tuneable lasers with appropriate sidebands and detuning, each molecule scatters approximately 400 photons during 2D laser cooling, limited by the interaction time and scattering rate. Leaks to dark states are less than 10$\%$. 
The experimental results are used to benchmark the trajectory simulations to predict the achievable flux $3.5 \ \unit{m}$ downstream for a planned $e$EDM experiment.
\end{abstract}

\maketitle

\section{\label{sec:Introduction}Introduction}
Intense beams of cold molecules provide valuable quantum sensors for testing fundamental symmetries and serve as the main loading mechanism of (magneto) optical traps used to study ultra-cold chemistry and quantum simulation \cite{ye2024essay-963,burau2023blue-detuned-e42, cheuk2020observation-8ca, blackmore2018ultracold-ba9, white2024slow-abd}. Heavy, polar molecules are particularly suitable for electron electric dipole moment ($e$EDM) searches of $\mathcal{CP}$ violation \cite{demille2024quantum-57b}. The current limit on the $e$EDM constrains some models of new physics up to and above 10 TeV, probing energies beyond those accessible by current particle colliders \cite{roussy2023improved-26e, cesarotti2019interpreting-4cb, athanasakis-kaklamanakis2025community-227}. Motivated by this, the field of molecular $e$EDM searches is rapidly developing \cite{hiramoto2023sipm-06c,  vutha2018oriented-b14, fitch2021methods-932}. This work is part of a project that aims at an $e$EDM measurement using an intense, cold, and slow beam of $^{138}$Ba$^{19}$F, henceforth BaF, molecules \cite{aggarwal2018measuring-5f1}, and to extend the methods used to trapped polyatomic molecules \cite{aggarwal2021deceleration-301, bause2025prospects-e91}. Furthermore, BaF is an attractive test case for experiments with isoelectronic, similarly structured but heavier radioactive molecules such as RaF~\cite{arrowsmith-kron2024opportunities-2fd} or RaOH~\cite{kozyryev2017precision-e8c}, which promise impressive improvements in sensitivity to nuclear $\mathcal{P,T}$ violating physics.

Current $e$EDM experiments use ions in a radiofrequency (RF) trap or neutral molecules in a cryogenic buffer gas beam (CBGB). The former yields a long coherent interaction time $\tau$, at the expense of smaller numbers of molecules $N$ due to Coulomb repulsion, e.g. $N\sim 2 \times 10^4$ HfF$^+$ ions with $\tau \sim 3 \ \unit{s}$ \cite{zhou2020second-scale-fbb, roussy2023improved-26e}. In contrast, the latter benefits from high molecule fluxes up to $\sim 10^{12}\ \unit{molecules \ state^{-1}sr^{-1}pulse^{-1}}$ but produces only $\tau \sim 1 \ \unit{ms}$ due to the beam's forward velocity and divergence \cite{andreev2018improved-575,hutzler2012buffer-3b9, truppe2018buffer-632}. 
A considerable fraction of the flux is generally lost due to the large spread of transverse velocity of molecules exiting the CBGB of $\sim 40 \ \unit{m/s}$. This loss can be reduced by implementing an electric or magnetic lens to capture and focus or collimate divergent molecular trajectories through a region of interest \cite{wu2022electrostatic-9bd, touwen2024manipulating-17c}. Other methods to reduce the transverse velocity are 'collimating' apertures to filter out transverse velocities above a few $\unit{m/s}$ \cite{kozyryev2017sisyphus-461, augenbraun2020laser-cooled-390}, or the application of Sisyphus laser cooling \cite{alauze2021ultracold-0d1}. Both of these methods select only a small part of the transverse phase-space distribution of the beam. In contrast, Doppler laser cooling can work on an arbitrary velocity range, set by the laser detuning. Therefore, an attractive option to obtain a collimated, high-flux molecular beam is to capture and focus a large part of the transverse-velocity distribution using an electrostatic lens and subsequently apply Doppler laser cooling near the focus. The combination of a hexapole lens and laser cooling enables the application of forces that depend on the transverse position and velocity of the molecules: The hexapole lens provides position-dependent forces, while laser cooling introduces velocity-dependent (friction) forces. The resulting intense beam of molecules can be used for precision tests directly, or as input for a Stark decelerator or longitudinal laser slowing, to improve loading rates into (magneto) optical traps \cite{langin2023toward-bb3}. 

Although various forms of laser cooling have been widely applied to atoms, extending these techniques to molecules is complicated due to their complex internal structure \cite{fitch2021laser-cooled-127}. However, in the last two decades, a variety of diatomic and polyatomic molecules have been cooled to low microkelvin temperatures and loaded into traps \cite{shuman2010laser-005, alauze2021ultracold-0d1, burau2023blue-detuned-e42, jorapur2024high-085}. Efficient transverse Doppler laser cooling of BaF is complicated by the relatively long lifetime of 57 ns of the A$^{2}\Pi _{1/2}$ state \cite{aggarwal2019lifetime-18a} and a small recoil velocity for each scattered photon due to the high molecular mass and low frequency of the cooling light. Furthermore, the hyperfine structure in both ground and excited states, the natural linewidth, and the relevant Doppler shifts are of similar magnitude. As a consequence, a single laser frequency will target different transitions and create competing blue-detuned heating and red-detuned cooling. Recent work on laser cooling BaF has demonstrated Sisyphus cooling in 1D \cite{rockenhuser2024laser-bb9}, and laser slowing and 3D magneto-optical trapping \cite{zeng2024three-dimensional-6be}. 

Here, we apply two-dimensional Doppler cooling on a hexapole-focused BaF molecular beam. We study the cooling effect as a function of laser detuning, laser power, and hexapole voltage and investigate its sensitivity to laser alignment. The experiment is modelled with trajectory simulations that use an effective two-level approach for the laser cooling effect. We used the simulations to extrapolate the gain in downstream molecule flux for various experimental configurations.

\begin{figure*}
\centering
\begin{minipage}{\textwidth}
\includegraphics[width=.99\columnwidth]{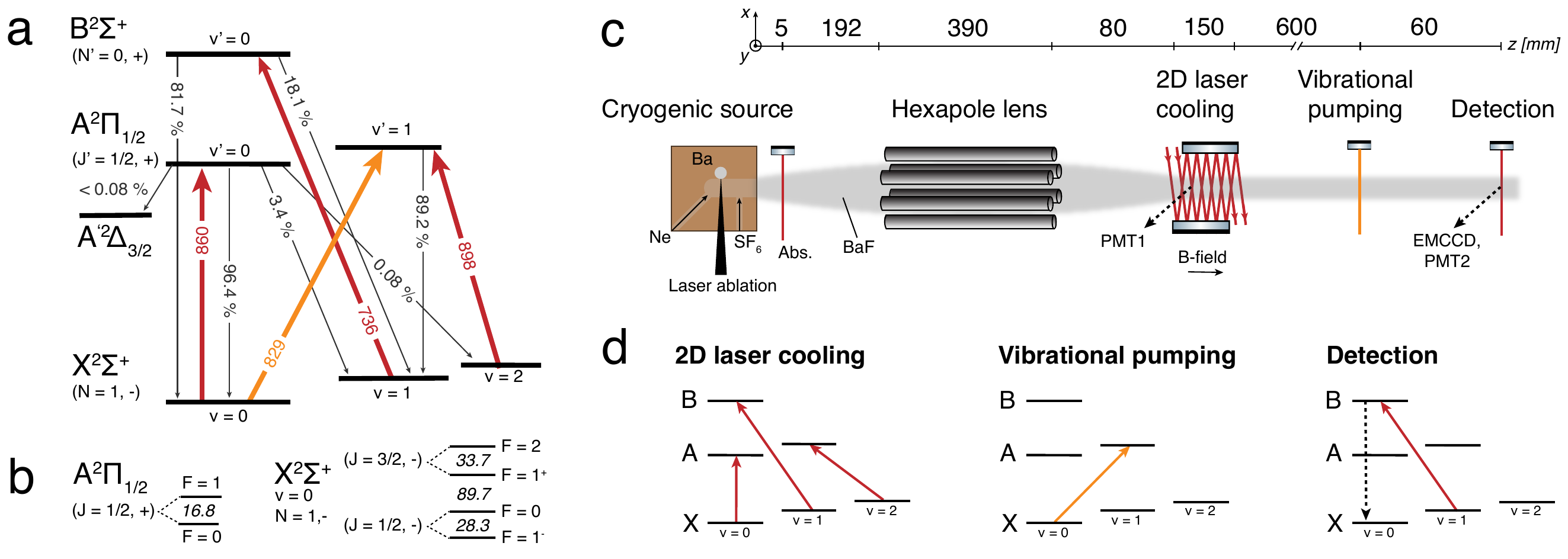}
\caption{(a) Energy level scheme relevant for laser cooling of $^{138}$Ba$^{19}$F. Colored lines indicate lasers used for cooling and vibrational (re)pumping, with corresponding wavelength in nm. Grey lines denote spontaneous decay channels with corresponding probabilities \cite{hao2019high-bd8}. (b) Hyperfine splittings in the X$^{2}\Sigma^{+} (v=0, N=1)$ and A$^{2}\Pi_{1/2} (v=0, J=1/2)$ state with level spacing in MHz. The B$^{2}\Sigma^{+} (v=0, N=0)$ hyperfine splitting is less than the natural linewidth of the transition. (c) Schematic view of the experimental setup and (d) laser light present in the experiment. The BaF molecular beam is created using a CBGB and focused into the laser cooling region using an electrostatic hexapole. There, the molecules interact with the main cooling light and two repump lasers (red). The reflection pattern shown here is present in both transverse directions. Next, the beam propagates through the vibrational pumping region, where molecules only experience vibrational depump light (orange). Lastly, an EMCCD camera records X $\leftarrow$ B fluorescence (dashed) after exciting molecules in the X $^{2}\Sigma^{+}(v=1, N=1)$ state.}
\label{fig:LevelScheme}
\end{minipage}\qquad
\end{figure*}
\section{\label{sec:Method}Method}
\subsection{\label{sec:ExpSetup}Experimental setup}
This section describes the laser system and the experimental setup. \autoref{fig:LevelScheme}a,b shows the relevant energy levels and hyperfine structure, including the branching ratios \cite{hao2019high-bd8}, for the laser cooling scheme applied here. We will refer to light that couples the electronic ground- and excited states with a specific vibrational level $v = i$ and $v' = j$, respectively, as $\mathcal{L}_{ij}$. We use in total three lasers (a cooling laser and two vibrational repumpers) with four sidebands each for the cooling process, a separate laser with one sideband for optical pumping into the detection state, and split off part of the repumpers to do efficient, background free detection. The main cooling laser $\mathcal{L}_{00}$ drives the rotationally closed X$^2 \Sigma^+ (v=0, N=1) \leftarrow$ A$^2 \Pi_{1/2} (v'=0, J'=1/2)$ transition with decay rate $\Gamma = 2\pi \times 2.79 \ \unit{MHz}$ and wavelength $\lambda = 860 \ \unit{nm}$ \cite{aggarwal2019lifetime-18a}. Dark Zeeman states are remixed using a magnetic field \cite{xia2021destabilization-ec2}. The total magnetic field is the sum of a $2 \ \unit{G}$ field in the $+z$ direction generated by a Helmholtz coil and the ambient magnetic field of $\sim 0.5 \ \unit{G}$, oriented mostly along $-y$. 
Molecules that decay to the vibrationally excited states $X (v = 1,2)$ are pumped back into the cooling cycle using lasers $\mathcal{L}_{10} \ \textnormal{and}\  \mathcal{L}_{21}$ at $736 \ \unit{nm} \ \textnormal{and} \ 898 \ \unit{nm}$, which drive the transitions X$^2 \Sigma^+ (v=1, N=1) \leftarrow$ B$^2 \Sigma^{+} \ ( v'=0, N'=0)$ and X$^2 \Sigma^+ (v=2, N=1) \leftarrow$ A$^2 \Pi_{1/2} (v'=1, J'=1/2)$, respectively. Each laser has RF sidebands added to address the relevant hyperfine structure. For $\mathcal{L}_{00}$, the $ F=2,1^{+},0,1^{-}$ levels are addressed using the first-order diffraction from four acousto-optic modulators (AOM) driven at $\{+77.0, +93.8, +200.4, +228.7\}$ MHz, respectively. $\sim 1 \ \unit{\micro W}$ of each $\mathcal{L}_{00}$ sideband is split off and used for absorption detection. For $\mathcal{L}_{10}$ and $\mathcal{L}_{21}$, the laser beam is sent through a resonant electro-optic modulator (EOM) which is driven using a sine wave with frequency $f_{m} = 38.5\ \unit{MHz}$ at, a modulation depth of $\phi_0 \sim 0.4 \pi$. This creates four symmetrically spaced sidebands that roughly overlap with the relevant transitions. The resulting sidebands are slightly off-resonant and consequently require some power broadening to excite the transitions, but this method is less complex compared to setting up a single AOM for each desired sideband, or generating the EOM sidebands using serrodyne waveforms \cite{kogel2025molecular-b3e}. 
The lasers are frequency stabilized using a HighFinesse WS8-2 wavemeter, which provides a short-term stability of $<0.5 \ \unit{MHz}$ but may drift several $\unit{MHz}$ on timescales of days.

$\mathcal{L}_{00}$, $\mathcal{L}_{10}$ and $\mathcal{L}_{21}$ are combined using polarizing optics and sent into two polarization maintaining fibers, one for each transverse cooling direction. $\mathcal{L}_{21}$ is only added to the horizontal cooling beam. At each fiber exit, the light is collimated to a $1/e^{2}$ waist of $3.3 \ \unit{mm}$ by $6.6 \ \unit{mm}$, parallel and orthogonal to the molecular beam propagation direction, respectively. The beam is then split into two parallel beams using a 50:50 non-polarizing beamsplitter cube, and aligned such that the reflections cross at the center of the molecular beam at an angle of $\approx 0.5\unit{\degree}$ with respect to orthogonal. A ghosting reflection of these two cubes is used to monitor the $\mathcal{L}_{00}$ laser power using a photodiode. Using a $\lambda/2$ plate, the angle of the linearly polarized cooling beams is set to $45\unit{\degree}$ with respect to $\hat{z}$, to maximize the rate at which dark $m_{F}$ states are remixed. The typical laser power per transverse direction delivered to the molecules for the wavelengths addressing X$ (v=0,1,2)$ is measured to be approximately $25,60,30 \ \unit{mW}$, respectively.

In order to facilitate detection, an additional laser $\mathcal{L}_{01}$ at 829 nm, referred to as the vibrational pump beam, is used to drive the X$^2 \Sigma^+ (v=0, N=1) \leftarrow$ A$^2 \Pi_{1/2} (v'=1, J'=1/2)$ transition. Since the A$^2 \Pi_{1/2} (v'=1, J'=1/2)$ state has a $\approx 90 \%$ probability to decay to the X$^2 \Sigma^+ (v=1, N=1)$ state \cite{hao2019high-bd8}, laser $\mathcal{L}_{01}$ effectively pumps molecules into the X$^2 \Sigma^+ (v=1, N=1)$ state. Molecules in this state are detected background-free  using $\mathcal{L}_{10}$ light. The 829 nm light passes through an AOM driven at $69 \ \unit{MHz}$. The generated zeroth-order and double-passed first-order beams are combined and used to create two equal intensity sidebands which address the $J=3/2$ and $J=1/2$ part of the X$^2 \Sigma^+ (v=0, N=1) \leftarrow$ A$^2 \Pi_{1/2} (v'=1, J'=1/2)$ transition, respectively. 

\autoref{fig:LevelScheme} shows the experimental setup (c) and the laser beams present at the relevant stages of the experiment (d). The experiment is located in a vacuum system with a pressure of $\approx 10^{-7} \ \unit{mbar}$. The cryogenic buffer gas beam source is identical to the one described in \cite{touwen2024manipulating-17c} except that the cell is extended from $11 \ \unit{mm}$ to $21 \ \unit{mm}$ to reduce the forward velocity \cite{mooij2024influence-779}. The source creates pulses of BaF molecules at 10 Hz with a mean forward velocity of $v_{z} = 184 \ \unit{m/s}$, with a variation of up to 10$\%$ depending on the operating parameters. Molecules exit the cell through an aperture with diameter $4.5 \ \unit{mm}$ at $z = 0 \ \unit{mm}$ and cross the double-passed absorption beam at $z = 5 \ \unit{mm}$. At $z = 187 \ \unit{mm}$, the molecules enter a hexapole lens with a length of $390 \ \unit{mm}$ and an inner diameter $d = 12 \ \unit{mm}$. The location of the hexapole focus can be changed by varying the applied voltage $V_{0}$. The transverse velocity-acceptance of the hexapole is equal to $v_{c} = 4.6 \ \unit{m/s}$ \cite{touwen2024manipulating-17c}, which limits the maximum transverse velocity of molecules that reach the laser cooling region. 

The laser cooling region starts $80 \ \unit{mm}$ after the hexapole. For both transverse directions, we introduce two parallel beams from one side of the laser cooling vacuum chamber which are retroreflected in such a way that the reflection of the first beam crosses the second beam as shown in \autoref{fig:LevelScheme}c. This ensures that at every crossing there is red-detuned light present for both positive and negative transverse velocities. In total, 42 (38) passes of cooling light are visible on the horizontal (vertical) vacuum chamber window, for a total effective interaction length of 145 (136) $\unit{mm}$. A photomultiplier tube (PMT1) is mounted at an angle of $45\unit{\degree}$ relative to the laser beams above the cooling region and collects fluorescence originating from $\approx 40 \ \unit{mm}$ after the start of the cooling region. 

The detection region is located $660 \ \unit{mm}$ downstream from the end of the laser cooling region. Here, molecules interact with the vibrational pump- and detection laser sheets, which both propagate along the $\hat{x}$ direction. Fluorescence is collected using a second photomultiplier tube (PMT2) mounted above the beam or an EMCCD camera mounted in the horizontal plane at a $\approx 55\unit{\degree}$ angle with respect to the molecular beam. The vibrational pump- and fluorescence detection beam have a $1/e^{2}$ waist of $\sim 1.5 \ \unit{mm}$ in the $z$ direction and are stretched in the $\hat{y}$ direction using cylindrical lens telescopes with magnification factors $M = 30$ and $M = 68$, respectively. The fluorescence beam is clipped in the vertical direction by the 1 inch optics, producing a sheet that is uniform over $\approx 20 \ \unit{mm}$ in the $y$ direction to within 75 $\%$ of its peak intensity. The vibrational pump- and fluorescence detection sheets have powers of $>150 \ \unit{mW}$ and $>50\ \unit{mW}$, respectively, and are retroreflected using a zero degree mirror. In order to compensate for the angle of the camera, the images are horizontally stretched during data analysis. The absorption signal is used to normalize all EMCCD and PMT fluorescence measurements, to cancel source fluctuations.  

\subsection{\label{sec:Trajectory_sims}Trajectory simulations}

Before presenting the experimental results, we will first describe the model used to simulate the cooling force. The molecular trajectory simulations presented in \cite{touwen2024manipulating-17c} are extended to account for changes in the cell geometry and incorporate the laser cooling section. The initial beam distribution is modelled by a 2D Gaussian distribution with a mean forward velocity of $v_z = 184 \ \unit{m/s}$, and a velocity spread of $30 \ \unit{m/s}$ and $40 \ \unit{m/s}$, in the longitudinal and transverse direction, respectively. The initial transverse position standard deviation is set to $4.5 \ \unit{mm}$. This is larger than the cell aperture and accounts for collisions in the beam occurring in the first few centimeters behind the cell. After this distance, collisions become infrequent, and the molecules follow ballistic trajectories. 

\begin{figure}
\centering
\includegraphics[width=0.99\columnwidth]{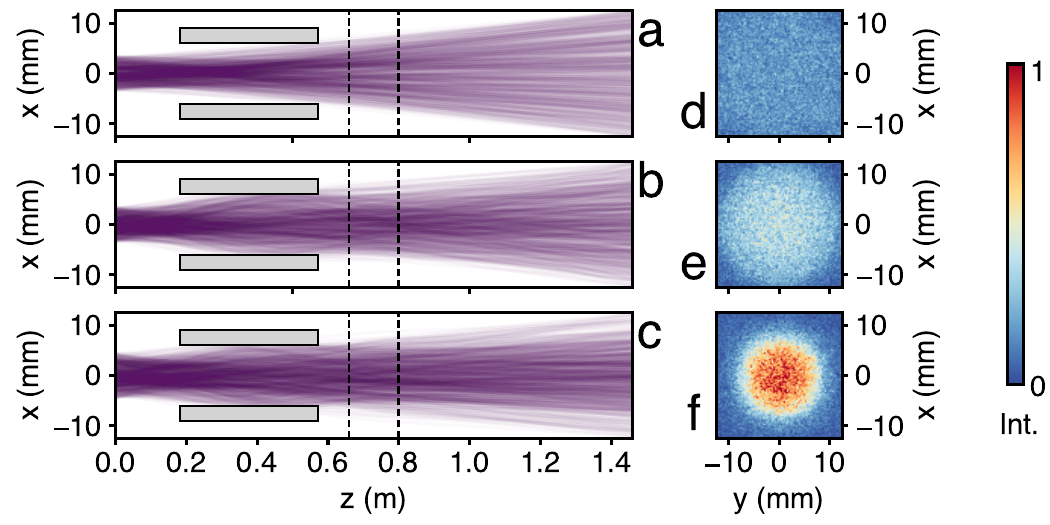}
\caption{Cross section in the $xz$ plane at $y=0$ showing simulated trajectories (a-c) along with transverse molecule distribution at the detection point (d-f), for three different situations. The position of the hexapole is indicated by the grey rectangles. The dashed lines indicate the laser cooling region. (a,d) No hexapole focusing or laser cooling. (b,e) Lensing at $2.0 \ \unit{kV}$ creates a semi-collimated beam. (c,f) Hexapole focusing with an applied voltage of $2.0 \ \unit{kV}$ and 2D laser cooling using $\Delta = - \Gamma, s_{0} = 4, \alpha = 0.5\unit{\degree}, \beta_{x},\beta_{y} = 0.05\unit{\degree}$ for each cooling direction. Here, we use $\eta = 0.125$ since the cooling beams overlap in the simulations, which will equally divide the total scattering rate between both directions.   
}
\label{fig:Trajectory_simulations}
\end{figure}

The Doppler force is modelled using a two-level approximation. The total acceleration $a$ is a sum of the force applied by two counterpropagating transverse laser beams: 
\begin{equation}
    a = \frac{\eta}{M} (F_{+} + F_{-}) 
\label{eq:friction}    
\end{equation}
where the mass of the BaF molecule $M = 156.325\ \unit{u}$. A factor $\eta$ is added, referred to as the scattering efficiency, to take into account the reduction of the scattering rate with respect to the two-level equivalent, e.g. due to the multilevel structure of the cooling transition. For a two-level system with ground state $g$ and excited state $e$, the maximum scattering rate $R_{sc} = \Gamma/2$. Here, $\Gamma = 1/\tau$, where $\tau$ and $\Gamma$ correspond to the lifetime and emission rate of the excited state, respectively. For a system with $N_{e} = 4$ excited states and $N_{g} = 12$ ground states, $R_{sc} = 2\Gamma N_{e} / (N_{e} + N_{g}) = \Gamma/4$ \cite{truppe2017molecules-6fa}. This results in a maximum total scattering efficiency of $\eta_{max} = 0.5$. Generally, the scattering rates measured in experiments with laser cooled molecules are significantly lower than their theoretical limit \cite{fitch2021laser-cooled-127}. In \autoref{sec:Scattering_rate}, we discuss how our derived scattering rates compare to the theoretical limit.   
The friction force of \autoref{eq:friction} will cool the molecules to arbitrarily low temperatures. To avoid this unphysical effect, the force is set to zero once the molecules reach a velocity below $v_{x,y} = \sqrt{k_{B}T_{D}/M} = 6 \ \unit{cm/s}$, with $T_{D} = \hbar \Gamma / 2k_{B}$ the Doppler temperature of the cooling transition $T_{D} = \hbar \Gamma / 2k_{B}$. 
In order to create the extended light field for cooling, the laser beams are reflected multiple times between two mirrors that are parallel to the molecular beam. As a result, the k-vector of the cooling light is not exactly orthogonal to the molecular beam axis but differs by a small angle $\alpha$. A nonzero $\alpha$ shifts the force curves resulting from each beam to higher velocities, which is compensated by reducing the detuning. Furthermore, an angle $\beta$ is introduced to account for a possible misalignment of the mirrors. The loss of laser power due to window reflections and mirror transmission exponentially reduces the effective saturation factor along the cooling region by a maximum of $50\%$, which is taken into account in the spatially dependent saturation parameter $s(z)$.  The force and detuning for both cooling beams present in the $x$ direction are thus defined as:\\
\begin{align}
    F_{+} &= \hbar k  \ \cos(\alpha - \beta)  \frac{\Gamma}{2}  \frac{s(z)}{1 + s(z) + \delta_{+}^{2}}, \label{eq:F+} \\  
    F_{-} &= \hbar k  \ \cos(\alpha + \beta)  \frac{\Gamma}{2}  \frac{s(z)}{1 + s(z) + \delta_{-}^{2}},\label{eq:F-}\\
    \delta_{+} &= \frac{2}{\Gamma} \big( (\Delta\Gamma) -k [v_{x}\cos(\alpha - \beta) + v_{z} \sin(\alpha - \beta)] \big), \label{eq:delta+}\\
    \delta_{-} &= -\frac{2}{\Gamma} \big( (\Delta\Gamma) +k [v_{x}\cos(\alpha + \beta) - v_{z} \sin(\alpha + \beta)]  \big)  \label{eq:delta-}
\end{align} 
The force and detuning in the $\hat{y}$ direction are defined using the same equations, substituting $v_{y}$ for $v_{x}$. The force model currently does not take into account the decay to dark states such as X$(v=3)$ or A$^{'2}\Delta_{3/2}$. These losses are taken into account by shifting the simulated detuning scan down so that it overlaps near $\Delta = 0$ with the measured signal. An example of a trajectory simulation for three different situations is shown in \autoref{fig:Trajectory_simulations}. Here, the value $\eta = \eta_{max}/2$ is used to determine the maximum signal gain expected in the experiment.
\section{Results}
\subsection{\label{sec:Spectroscopy}Spectroscopy}
\begin{figure}
\includegraphics[width=.8\columnwidth]{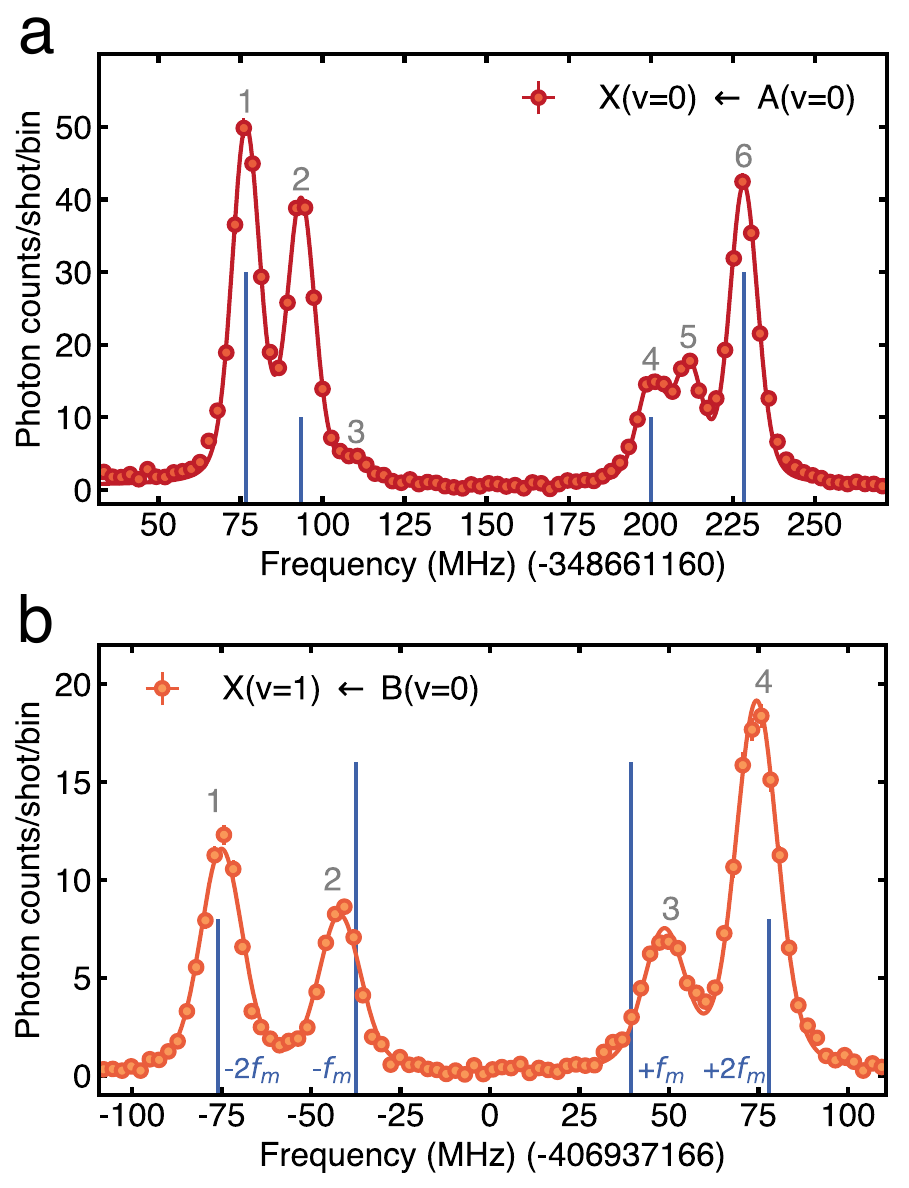}
\caption{Spectroscopy of the main cooling transition (a) and the first vibrational repump transition (b) recorded with PMT1. The sidebands (blue) are set at frequencies corresponding to certain $F \rightarrow F'$ transitions (grey), which are extracted from a multiple Voigt profile fit (solid line). The sideband heights reflect their power ratio. (a) The  X$^2 \Sigma^+ (v=0, N=1) \leftarrow$ A$^2 \Pi_{1/2} (v=0, J=1/2)$ transition. (b) The X$^2 \Sigma^+ (v=1, N=1) \leftarrow$ B$^2 \Sigma_{1/2} (v=0, N=0)$ transition. The excited state hyperfine splittings is not resolved. }
\label{fig:X0A0_X1B0_spectra}
\end{figure}

\begin{figure}
\includegraphics[width=0.8\columnwidth]{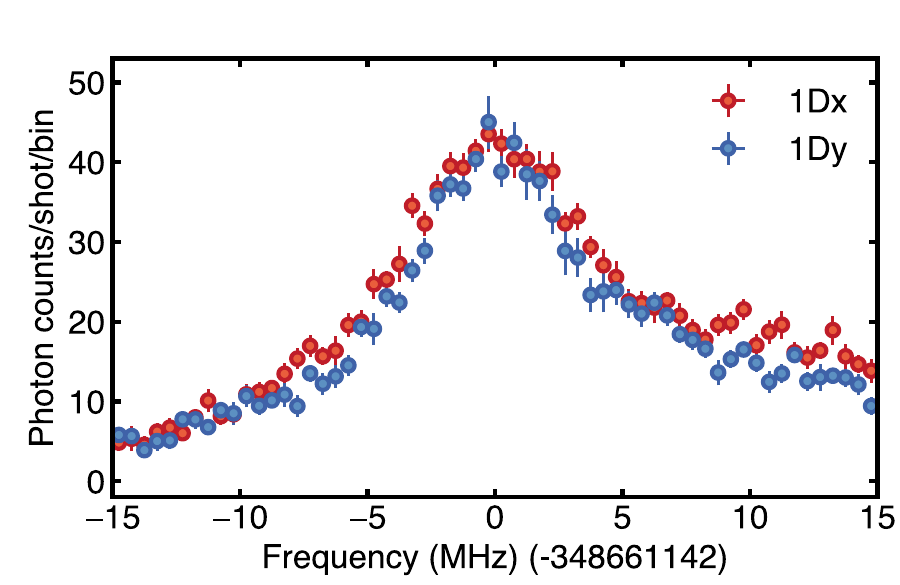}
\caption{Spectroscopy the X$(v=0) \rightarrow$ A$(v=0)$ transition for horizontal (1Dx) and vertical (1Dy) laser cooling. The spectrum shown here is obtained by varying the laser frequency while the RF sideband spacings are kept constant. At a frequency measured by the wavemeter of $ 348661142.0(5) \ \unit{MHz}$, the AOM sidebands are resonant with transitions 1, 2, 4 and 6 from \autoref{fig:X0A0_X1B0_spectra}a, which gives the frequency corresponding to $\Delta = 0 \ \unit{MHz}$. The $\approx 20 \ \unit{MHz}$ offset compared to \autoref{fig:X0A0_X1B0_spectra}a is due to long term drift of the wavemeter.}
\label{fig:L00_spectrum}
\end{figure}

In order to determine the required RF sideband frequencies of all lasers present in the cooling region, we performed spectroscopy with a single frequency beam using a laser power such that $s<1$ on the transition addressed by each laser. The resulting spectra are shown in \autoref{fig:X0A0_X1B0_spectra}, including the relative frequency spacing and laser power of the RF sidebands. For (a), the $+77 \ \unit{MHz}$ sideband was used, and for (b) the EOM was simply disabled. Using these spectra, the power ratio for the $\mathcal{L}_{00}$ sidebands was chosen such that the strongest transitions are addressed by the highest laser power while minimizing off resonant excitation from a sideband that targets a different transition. For $\mathcal{L}_{10}, \mathcal{L}_{21}$, the power ratio of the symmetrically spaced sidebands is set to $\approx 2:1$ to compensate for the fact that the first-order sidebands are further detuned from resonance than the second-order. Laser $\mathcal{L}_{21}$ uses the same sideband structure as laser $\mathcal{L}_{10}$. In all experiments, except for the one shown in \autoref{fig:X0A0_X1B0_spectra}a, narrow bandpass filters with a transmission wavelength of 712 nm are placed in front of the PMTs and EMCCD camera in order to detect fluorescence from the X$(v=0) \leftarrow$ B$(v'=0)$  transition, while blocking out stray light.
Before taking a laser cooling data set, we record the spectrum of the main cooling transition to confirm that the angular frequency corresponding to a detuning of $\Delta = 0 \ \unit{MHz}$ is equal to within $1 \ \unit{MHz}$ for both directions, which we take as evidence that the cooling light is correctly aligned. The RF-frequencies to the AOMs are kept constant while the input laser frequency is varied. \autoref{fig:L00_spectrum} shows such a spectrum, obtained when $\mathcal{L}_{00}$ is reduced to $\approx 2 \unit{mW}$. The observed asymmetric profile results from the relative strengths of the different hyperfine components shown in \autoref{fig:X0A0_X1B0_spectra}a.  

Next, a spectrum is recorded using PMT2 for lasers $\mathcal{L}_{10}$ and $\mathcal{L}_{21}$ and they are locked at the frequency that produces maximum fluorescence for an optimal repumping rate. The signal from PMT2 rather than PMT1 is used to determine these resonances to minimize the detection bias due to the Doppler shifts from the horizontal velocity component of the molecules. The mean beam velocity is calculated from the mean arrival times measured by PMT1 and PMT2. Typically, after locking the lasers, images of the 712 nm fluorescence detected on the EMCCD camera are averaged over 600 shots, where for each shot a background image is subtracted.

\subsection{\label{sec:Scattering_rate}Scattering rate}
\begin{figure}
\centering
\includegraphics[width=.912\columnwidth]{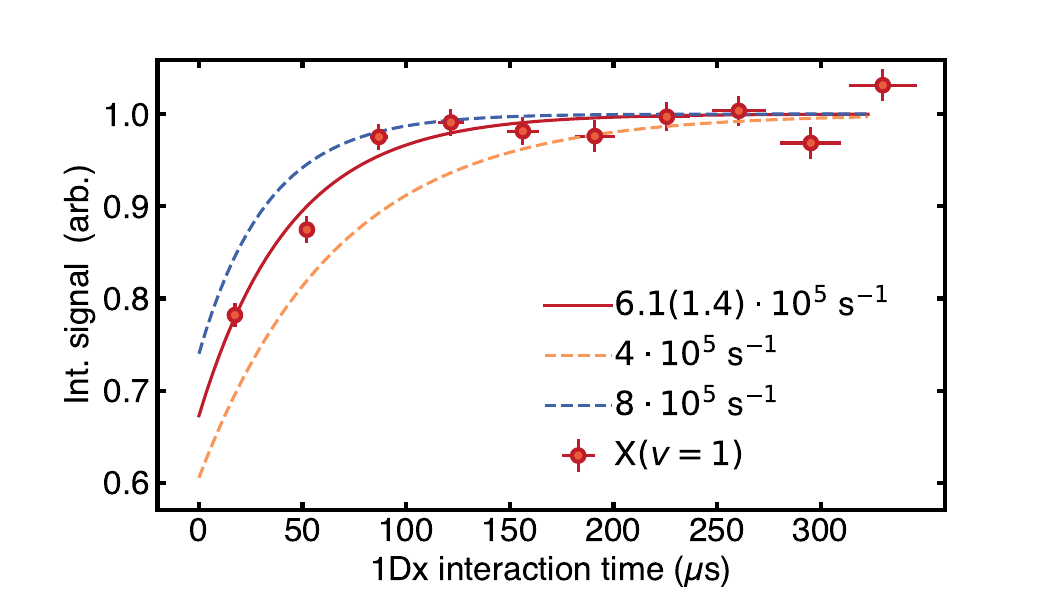}
\caption{Integrated camera signal versus interaction time with crossings of horizontal cooling light shows cycling into the X$(v=1)$ state. The only lasers present here are 1Dx $\mathcal{L}_{00}$ in the cooling region and $\mathcal{L}_{10}$ at the detection point. The fitted scattering rate (red) is shown, along with a lower (yellow) and higher (blue) scattering rate for illustration purposes. The signal starts at a nonzero value due to background X$(v=1)$ population and cycling into the X$(v=1)$ from two passes of the cooling light that do not cross another beam. }
\label{fig:Rscat_fit}
\end{figure}

To estimate $\eta$, the scattering rate of the main cooling transition is measured by detecting the population in the X$(v=1)$ state using $\mathcal{L}_{10}$ and while cycling with only $\mathcal{L}_{00}$ light and all repumpers off. \autoref{fig:Rscat_fit} shows the full image EMCCD signal as a function of interaction time with only horizontal cooling light (1Dx), including a fit for the scattering rate, the signal at $t=0 \ \unit{\micro s}$ and $t = 340\ \unit{\micro s}$. The number of crossings, where red-detuned light is present for both positive and negative transverse velocities, is varied by moving a beam block in front of one of the cooling mirrors. The number of crossings is converted to an interaction time using the measured $1/e^{2}$ width of the cooling beam and the mean forward velocity derived from the time of flight profiles recorded by PMT1 and PMT2. The fitted signal at $t=0$ corresponds to background X$(v=1)$ population. The fit uses the relative branching fraction of A$^{2}\Pi_{1/2} (v=0) \rightarrow$ X$^{2}\Sigma^{+} (v=0)$, calculated to be $96.4\%$ \cite{hao2019high-bd8}. From this fit, a scattering rate of $R_{sc} = 6.1(1.4)\times10^{5}\ \unit{s^{-1}}$ is extracted, which is $14 \%$ of the theoretical maximum of $\Gamma/4 = 4.4 \times 10^{6} \ \ \unit{s^{-1}}$. The actual scattering rate during laser cooling is expected to be lower as vibrationally excited molecules are not repumped immediately. We assume a constant scattering rate for both 1D and 2D cooling, which is implemented by introducing an offset between the reflections, such that a low intensity region for 1Dx overlaps with a high intensity region for 1Dy and vice versa. 

This scattering rate may be compared to those obtained in other experiments. Previous work with BaF showed a scattering rate approaching $80\%$ of the theoretical maximum \cite{chen2017radiative-cc3}. More recently, a scattering rate of $7(1)\times 10^{5} \ \unit{s^{-1}}$ was measured for BaF trapped in a MOT \cite{zeng2024three-dimensional-6be}. Although the latter is similar to our value, the maximum scattering rate in that experiment is reduced due to the presence of microwaves coupling the X$(v=0, N=0)$ and X$(v=0, N=1)$, increasing the number of ground states of the cooling transition. These values suggest that with a better controlled magnetic field, our measured scattering rate can still be increased, which would improve the cooling effect. 
\subsection{\label{sec:LC_results}Laser cooling characterization}
\begin{figure}
\centering
\includegraphics[width=0.99\columnwidth]{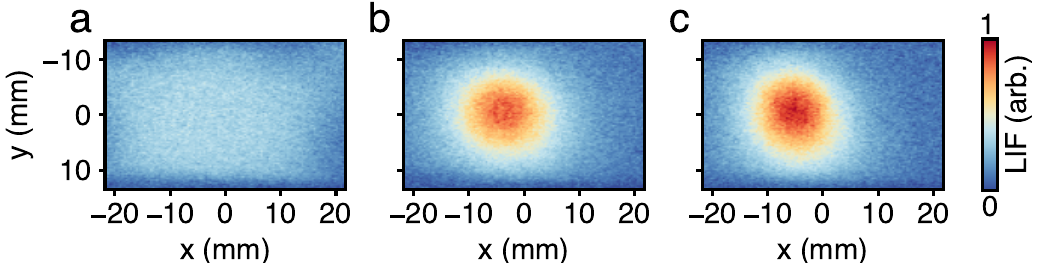}
\caption{Measured laser induced fluorescence (LIF) of the hexapole focused and 2D laser cooled molecular beam. With both the hexapole lens and laser cooling disabled, molecules are spread out uniformly over the image (a). A faint shadow from to the hexapole electrodes is visible. The sharp decrease in signal in the top and bottom of the image is due to the size of the detection light. With the hexapole lens set at $2.0 \ \unit{kV}$, a semi-collimated beam of low field seeking molecules is created, while high field seekers are defocused (b). With the hexapole lens at 2.0 kV and with 2D laser cooling, the transverse velocity is further reduced and the peak signal increases (c).}
\label{fig:Sketch_nocooling_vs_bestcooling}
\end{figure}

This section details the measurement procedure to quantify the effect of laser cooling and presents the main results. \autoref{fig:Sketch_nocooling_vs_bestcooling}a-c shows three example images for different experimental realizations. In (a), both the hexapole lens and laser cooling are turned off, which produces a uniform molecule density, clipped only by the hexapole electrodes and the detection beam. Since the detection beam does not cover the full image height and the CCD chip is less sensitive near the edges, further signal processing is generally restricted to a rectangle of $36 \ \unit{mm}$ in $\hat{x}$ by $18\ \unit{mm}$ in $\hat{y}$, referred to as the "full image". With the lens set to $2.0 \ \unit{kV}$, a semi-collimated beam is created (b). We chose a hexapole voltage of $2.0 \ \unit{kV}$ in these measurements because that gave the clearest and most easy-to-interpret demonstration of laser cooling. For optimal flux far from the source, the optimal hexapole voltage is somewhat higher, such that the molecules are focused in the laser cooling region and subsequently rapidly cooled, as shown in \autoref{sec:implications_for_NL-eEDM}. 
Finally, when cooling light with detuning $\Delta = - 2.5(5) \ \unit{MHz}$ is added in both horizontal and vertical directions (c), the peak intensity of the beam further increases, indicating a reduction in transverse velocity due to laser cooling. 

\begin{figure*}
\centering
\begin{minipage}{\textwidth}
\includegraphics[width=.99\columnwidth]{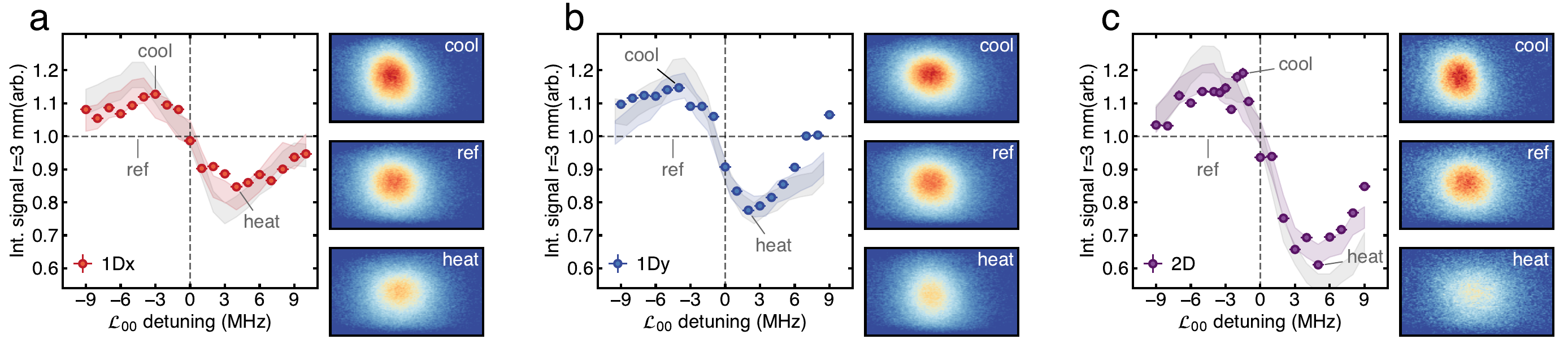}
\caption{Measured (datapoints) and simulated (shaded) signal intensity versus cooling laser detuning, for one dimensional transverse cooling along $\hat{x}$ (a), along $\hat{y}$ (b) and transverse cooling in two dimensions simultaneously (c). Error bars in signal and frequency denote $1\sigma$ uncertainty and 0.5 MHz lock uncertainty, respectively. Two simulations are included in each panel; one with a value for the scattering efficiency as determined by the scattering rate measurement shown in \autoref{fig:Rscat_fit} (grey), and one with a 20$\%$ reduced scattering efficiency (color). The insets next to each curve show an image with negative (positive) detuning labeled 'cool' ('heat'), with the corresponding frequency indicated in the main plot. The middle image is a reference image ('ref') without $\mathcal{L}_{00}$ light applied, but with $\mathcal{L}_{10}$ and $\mathcal{L}_{21}$ present. The image dimensions and color scale of the insets are identical to those of \autoref{fig:Sketch_nocooling_vs_bestcooling}. $\mathcal{L}_{00}$ laser power used here was $29 \ \unit{mW}$ for 1Dx and $25 \ \unit{mW}$ for 1Dy.}
\label{fig:L00freqscan}
\end{minipage}\qquad
\end{figure*}

\autoref{fig:L00freqscan} shows a frequency scan in a range of approximately $[-3\Gamma, +3 \Gamma]$ using only horizontal cooling light (a, 1Dx), only vertical cooling light (b, 1Dy) or both (c, 2D). The recorded 2D images are analyzed as follows; First, 1D Gaussians of the form $A\cdot\textnormal{exp}[{(x-x_{0})^{2}/{2\sigma_{x}^{2}}}]$ and $A\cdot\textnormal{exp}[{(y-y_{0})^{2}/{2\sigma_{y}^{2}}}]$ are fitted to the histogram of the signal along $x$ and $y$, respectively. Next, a circle centered on $x_{0},y_{0}$ with radius $r = 3 \ \unit{mm}$ is drawn. The number of counts inside this circle is normalised to a reference image (ref) recorded without $\mathcal{L}_{00}$ light present in the cooling region and depicted as the red, blue and purple circles in (a,b,c). The shaded curves, also shown, result from trajectory simulations aimed at replicating the experimental conditions; one with the value for the scattering efficiency determined by the scattering rate measurement shown in \autoref{fig:Rscat_fit} (grey), and one with a 20$\%$ reduced scattering efficiency (color) attributed to the fact that molecules spent some time in excited vibrational states before being pumped back into the cooling cycle. Multiplying this reduced scattering rate with the total interaction time of $\approx 0.8 \ \unit{ms}$ yields a total number of scattered photons of $N_{\gamma} = 390^{+90}_{-80}$. The image labeled cool (heat) shows the recorded image at the negative (positive) detuning indicated by the arrow. The dispersive shapes showing an increase (decrease) in signal at negative (positive) detuning are indicative of Doppler cooling (heating). For small negative detunings, a $10-15\%$ and $20 \%$ signal increase is observed for 1D cooling and 2D cooling, respectively. The colored shaded curves for 1Dx, 1Dy and 2D cooling are shifted down by $3\%$, $1\%$ and $10\%$, respectively, which we attribute to losses to dark states. Furthermore, the difference in signal between the cooling and heating peak is extracted. This value, labeled $\Delta S_{pp}$, is a useful metric to estimate the scattering efficiency $\eta$ used in the trajectory simulations. For the data shown in \autoref{fig:L00freqscan}a,b, the extracted values of $\Delta S_{pp}$ are 0.28(2) and 0.37(2), respectively. When 2D cooling light is applied, $\Delta S_{pp}$ roughly doubles to 0.59(2), indicating the successful application of cooling or heating forces simultaneously in both transverse directions. 

\begin{figure}
\centering
\includegraphics[width=.99\columnwidth]{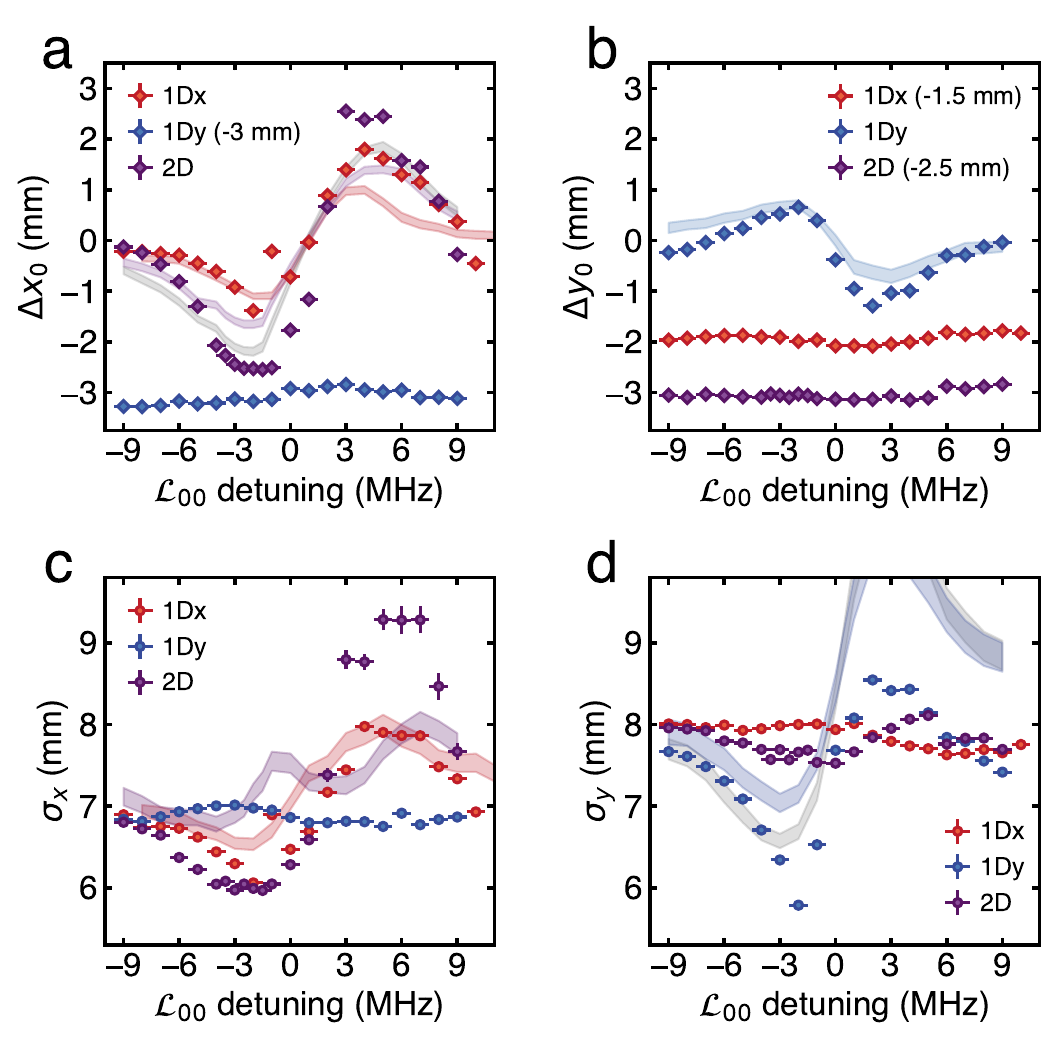}
\caption{Displacement of fitted center positions compared to that of the reference image (a,b) and fitted standard deviations (c,d) as a function of cooling laser detuning. Vertical error bars denote uncertainty in least-squares fit. The fluorescence imaging system does not capture the wings of the profile in the vertical direction, leading to uncertainties on $\sigma_{y}$ of $> 20\%$. Therefore, vertical error bars in (d) have been omitted for clarity. The 1Dy data in (a) and the 1Dx and 2D data in (b) is offset for clarity. }\label{fig:L00freq_sigma_centerpos}
\end{figure}

\autoref{fig:L00freq_sigma_centerpos} shows the displacement of the fitted center position $\Delta x_{0}, \Delta y_{0}$ (a,b) and the fitted standard deviation $\sigma_{x}, \sigma_{y}$ (c,d) as a function of laser frequency for the same measurements shown in \autoref{fig:L00freqscan}. The signal increase in \autoref{fig:L00freqscan} is attributed to a decrease in $\sigma_{x}$ and $ \sigma_{y}$. The center position of the molecular beam varies as a function of cooling laser detuning, which we attribute to a non-zero $\beta$, signalling that the mirrors are not perfectly parallel to the axis along which the molecular beam propagates ($\hat{z}$). The resonance frequency for each cooling beam changes due to the additional transverse velocity component in the mirror frame originating from the forward velocity $v_{z}$. As $v_{z}/v_{t} \approx 50$ in the cooling region, a value of $\beta = 1\unit{\degree}$ already shifts the force curves by $\sim3 \ \unit{m/s}$. Consequently, a non-zero $\beta$ leads to i) molecules being cooled to some nonzero transverse velocity and ii) the center position of the beam shifting as a function of laser detuning. 
The misalignment of the horizontal cooling mirrors was reduced between the 2D and 1Dx data sets, which explains the reduced increase in $\sigma_{x}$ at a heating frequency and the reduced peak-to-peak $\Delta x_{0}$ for 1Dx compared to 2D. 
Next, we simulate trajectories using various $\eta,\beta$ and take the combination that minimized the least squares residual between the measured and simulated displacement. We focus on displacement rather than standard deviation because the former is less sensitive to camera angle and possible velocity selectivity of the detection method. The shaded curves in \autoref{fig:L00freq_sigma_centerpos}a,b show the simulated displacement using values for $\eta,\beta$ that best match the displacement obtained experimentally. Using this procedure, we obtain misalignment angles of $\beta_{x} = 0.45 \unit{\degree}$ for 1Dx, $\beta_{y} = 0.2 \unit{\degree}$ for 1Dy and $\beta_{x} = 0.9 \unit{\degree}, \beta_{y} = <0.05 \unit{\degree}$ for 2D. The negligible value of $\Delta y_{0}$ using 2D cooling light suggests that the vertical cooling mirror was almost perfectly parallel to the molecular beam. Alternatively, it can be caused by a larger scattering rate in the horizontal compared to the vertical direction, which is currently not taken into account in the model.

We attribute the apparent $\approx 1 \ \unit{mm}$ offset in $\sigma_{x}$ versus $\sigma_{y}$ to a reduced detection efficiency of molecules with higher horizontal velocities. The simulations show that these are on average located further from the center of the image, as opposed to molecules with lower transverse velocities, which are generally found in the center region of the beam. To model the reduced detection efficiency of molecules that are Doppler-shifted out of resonance due to their horizontal velocities, we record the horizontal velocity distribution of all molecules and mask it using a Lorentzian centered at $0 \ \unit{m/s}$ with a full width at half maximum (FWHM) between 0 and $10 \ \unit{m/s}$. We find that a FWHM of $6 \ \unit{m/s}$ minimizes the least-squares residuals between the measured and simulated signal in the full image using 1Dx cooling, suggesting a saturation of $s\approx 1$ for the pump and detection transitions. This detection bias originates from the fact that $\mathcal{L}_{01}$ or $\mathcal{L}_{10}$ are insufficiently intense. $\mathcal{L}_{01}$ has only two sidebands compared to the four sidebands of $\mathcal{L}_{10}$. While the $\mathcal{L}_{01}$ pump light has a factor $\sim 3$ more power compared to the $\mathcal{L}_{10}$ detection light, the branching ratio of the X$(v=0) \rightarrow$ A$(v=1)$ transition is a factor $\sim 7$ larger compared to the X$(v=1) \rightarrow$ B$(v=0)$ transition \cite{hao2019high-bd8} driven by the laser $\mathcal{L}_{10}$, which leads to a higher saturation intensity and consequently to a reduced amount of power broadening. Therefore, the velocity selection is probably due to the $\mathcal{L}_{01}$ beam. 

The laser cooling section reduces the spread of the transverse velocity of the molecular beam, resulting in a decrease (increase) in $\sigma_{x}, \sigma_{y}$ at negative (positive) detuning. For laser cooling in only one dimension, the randomly directed emission pattern of Doppler cooling creates a small heating effect in the direction orthogonal to the plane spanned by the cooling light. The values obtained previously for $\eta, \beta$ are used for the shaded curves in (c,d). Similarly to \autoref{fig:L00freqscan}, a grey curve uses a reduced scattering efficiency determined by the measurement of the scattering rate shown in \autoref{fig:Rscat_fit} and the colored curve uses a 20$\%$ reduced scattering efficiency. The shaded curve for 2D cooling in (c) shows an interesting feature near 0 detuning, where a heating (cooling) effect appears at small negative (positive) detuning. We attribute this to the fact that when $\beta > \alpha$, the Doppler shift due to the forward velocity of the beam has opposite sign for both beams in one direction. However, the experimental data does not show this feature and therefore suggests instead that the observed displacement is due to a higher scattering rate and a smaller angle $\beta_{x}$. The shaded curves in (d) are not subject to the horizontal velocity detection and confirm that the standard deviation of the beam in the absence of laser cooling is $\approx 8 \ \unit{mm}$, but we leave out a more quantitative comparison to the experiment due to the large uncertainty on the fit to the experimental data. 

\begin{figure}
\centering
\includegraphics[width=.99\columnwidth]{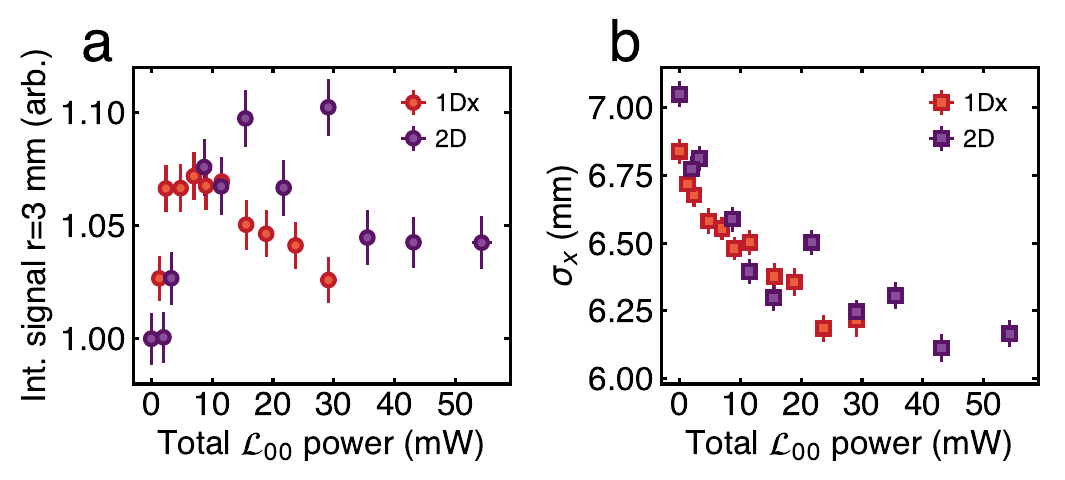}
\caption{Measured signal intensity (a) and fitted standard deviation (b) as a function of cooling laser power, for 1Dx cooling (red) and 2D cooling (purple). }
\label{fig:EMCCD_L00powerscan}
\end{figure}

\autoref{fig:EMCCD_L00powerscan} shows the measured signal in a $r=3 \ \unit{mm}$ circle (a) and the fitted $\sigma_{x}$ (b) as a function of laser power. Here, the laser detuning was set to $\Delta = -2.5(5) \ \unit{MHz}$. In panel (a), the signal sharply rises at low powers for both 1D and 2D cooling, before flattening off and eventually declining. However, in panel (b), we show that the width of the beam only starts to saturate at the highest measured laser powers. This likely indicates that at higher laser powers, the scattering rate and consequently the cooling effect increase, but the loss to dark states counteracts the signal increase. We expect the cooling to improve when we add repumpers to address the leaks to the positive parity X$^{2}\Sigma (v=0, N=0,2)$ states, which originate from decay to the intermediate electronic A$^{'2}\Delta _{3/2}$ state, or decays directly to the vibrationally excited X$^{2}\Sigma^{+} (v=3, N=1)$ state.  

\section{\label{sec:implications_for_NL-eEDM}Impact on NL-eEDM experiment}
\begin{figure}
\centering
\includegraphics[width=.99\columnwidth]{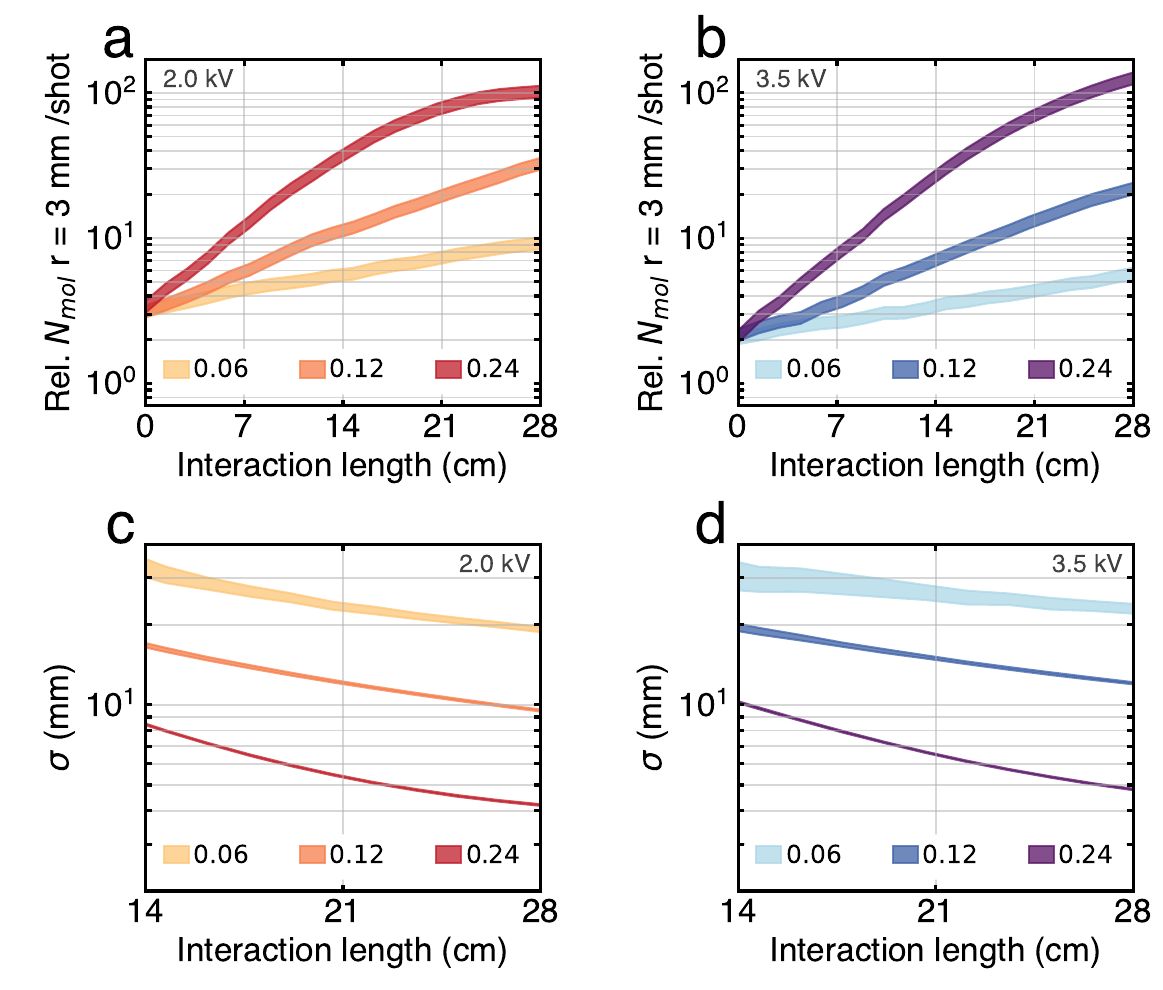}

\caption{Simulated relative molecule flux per shot (a,b) and beam size (c,d) for increasing laser cooling interaction length. The flux is normalized to the simulated flux with the hexapole voltages and cooling light turned off. Here, we use three different scattering efficiencies of $\eta = 0.06, 0.12, 0.24$, and a hexapole voltage of $2.0 \ \unit{kV}$ (yellow, orange, red) or $3.5 \ \unit{kV}$ (light blue, dark blue, purple). Installing two laser cooling vacuum chambers provides a total interaction length of $28 \ \unit{cm}$. The cooling parameters are $\Delta = - \Gamma, s = 4,\alpha = 0.5\unit{\degree}$ with angles of $ \beta_{x},\beta_{y} = 0.05\unit{\degree}$ for each cooling direction. }
\label{fig:sim_interaction_time_scan}
\end{figure}

We plan to implement the method to obtain a bright beam by combining a hexapole lens and laser cooling presented in the next phase of the NL-$e$EDM experiment to increase both the molecule flux and maximum coherent interaction time. The supersonic source used in the NL-$e$EDM experiment \cite{boeschoten2024spin-precession-17e} has a mean forward velocity of $\approx 600 \ \unit{m/s}$ and brightness of $6\times 10^{8}\ \unit{molecules/pulse/sr}$ in the X$^{2}\Sigma^{+} (v=0, N=1)$ state \cite{aggarwal2021supersonic-df1}. In contrast, the cryogenic buffer gas source we use has a mean forward velocity of $\approx 200 \ \unit{m/s}$ and an initial brightness of $\approx 2\times10^{10}\ \unit{molecules/pulse/sr}$ in the X$^{2}\Sigma^{+} (v=0, N=1)$ state \cite{mooij2025cryogenic-74c}, similar to values reported for other molecular species \cite{wright2023cryogenic-9e0}. The exact gain in molecule flux due to implementing the focused and laser cooled cryogenic beam depends on the efficiency of the hexapole lens-laser cooling combination, which in turn strongly depends on the voltage applied to the hexapole lens, and the attainable scattering rate and interaction time in the laser cooling process. The free flight distance for the beam in the NL-$e$EDM experiment is $\approx 3.5 \ \unit{m}$. \autoref{fig:sim_interaction_time_scan} shows the simulated relative molecule flux (a,b) in a circle of radius 3 mm centered on the molecular beam with $v_{z} = 184 \ \unit{m/s}$ at $z=\ 3.5\ \unit{m}$ downstream from the laser cooling chamber and beam size (c,d), for varying laser cooling lengths, scattering efficiencies and hexapole voltages, and is normalized to the flux with the hexapole voltages and laser cooling light turned off. Here, we use the same dimensions and spacing of the cooling light as shown in \autoref{sec:ExpSetup}, and assume the population to be entirely in a low-field seeking state. This can be experimentally achieved by using 3 out of 4 sidebands that address the main cooling transition, to pump molecules residing in hyperfine states $F=0,1^{-}, 1^{+}$ into the low-field seeking $F=2$ state. The simulations suggest that implementing a second laser cooling chamber and increasing the scattering rate to a realistic value of $50\%$ of the theoretical maximum will increase the flux of molecules in the $N=1$ state by more than a factor 100 compared to that of the cryogenic source. 
A separate manuscript with a more detailed discussion on the exact implementation of rotational- and hyperfine pumping stages, the lens and the laser cooling region, population transfer to the $e$EDM measurement state and the resulting relevant molecule flux for the NL-$e$EDM experiment is currently in preparation. 
\section{\label{sec:discussion}Discussion and conclusion}
The main factors that currently limit the efficiency of the laser cooling process are the scattering rate, the interaction time and the decay to dark states. We attribute the measured reduced scattering rate largely to, in order of importance, suboptimal dark Zeeman state remixing, sideband power ratios and detunings, and a lack of vibrational repump laser intensity. Improved control over the angle between laser polarization and magnetic field after compensating the background magnetic field in combination with a lower magnetic field strength is expected to produce a higher remixing rate. In the future, we plan to experimentally investigate whether the scattering rate can be further increased by changing the sideband power ratios and spacings, setting e.g. the $F=2$ sideband at $\Delta = - \Gamma$ and the $F=0$ sideband at $\Delta = -\Gamma/2$. Furthermore, our current trajectory simulations use a two-level approximation that does not take into account molecule-specific effects such as off-resonant excitation or coherent effects arising from the sideband structure. In the future, we plan to use PyLCP \cite{eckel2022pylcp-c00} to simulate the full optical Bloch equations for our system to reveal deeper insights into the cooling cycle and take sub-Doppler cooling effects into account.

The interaction time can be increased by extending the cooling region or reducing the forward velocity of the molecular beam. The former can be easily extended by adding a second cooling chamber, assuming that there is enough laser power present to sufficiently saturate the required transitions. The latter can be improved by installing a second cell after the cryogenic source \cite{white2024slow-abd}, which was shown to lower the forward velocity by $\approx 30\%$ without sacrifice of $N=1$ flux. The decay to dark states can be addressed by adding repumpers for the relevant states. 

Three different, more fundamental limits of the lens-laser cooling combination are the Stark shift of the lensing state, the capture velocity of the hexapole lens, and the rate of velocity dissipation due to the lifetime of the excited state of the cooling transition. The Stark shift of the lensing state $N=1$ of BaF has a quadratic or linear character depending on the applied electric field, which does not allow one to create a harmonic potential with only a single quadrupole or hexapole lens. This prevents the creation of a tight focus \cite{touwen2024manipulating-17c}. Furthermore, the turning point from low-field to high-field seeking is at a rather modest field strength of $\approx 20 \ \unit{kV/cm}$, which combined with the relatively high mass of the BaF molecule limits the capture velocity to $\approx 4.6 \ \unit{m/s}$, which implies that only a small fraction of the molecules that exit the CBGB are used. Alternatively, a higher rotational state such as $N=2$ can be used, which yields a higher lens capture velocity but requires state transfer and higher voltages. Therefore, the hexapole lens-laser cooling combination is likely better suited for lighter, laser-cooleable molecules with i) a higher maximum scattering rate and higher lens capture velocity, such as CaF or SrF or ii) a state possessing a fully linear or quadratic Stark shift, such as the (010) bending mode of CaOH or SrOH\cite{fitch2021laser-cooled-127}. If the hexapole lens capture velocity exceeds the laser cooling capture velocity, a curved wavefront approach can be used \cite{clausen2024combining-62c}, or the laser detuning can be swept in a manner similar to chirped laser slowing to ensure molecules stay resonant as their transverse velocity reduces \cite{truppe2017intense-65d}. The laser cooling section can also be split into a Doppler and sub-Doppler part, where the first section uses Doppler cooling to reduce the transverse velocity to a range where strong sub-Doppler forces can be leveraged to create an ultracold beam.

In conclusion, we have demonstrated and analyzed the combination of a cryogenic buffer gas source, an electrostatic hexapole lens and 2D transverse Doppler laser cooling to produce an intense, cold and collimated beam of $^{138}$Ba$^{19}$F molecules. We modelled the laser cooling region using an effective two-level approach and used it to explain the measured signal increase and molecular beam properties. We measure a scattering rate of the main cooling transition that is $14\%$ of the theoretical maximum and find that a further $\approx 20\%$ reduced rate is needed for the simulated cooling effect to match the experimental data. These simulations also predict a 100-fold increase in molecule flux available for our $e$EDM experiment. 


\section{\label{sec:Acknowledgements}Acknowledgements}
The NL-$e$EDM collaboration receives funding (eEDM-166, XL21.074 and VI.C.212.016) from the Dutch Research Council (NWO). We acknowledge the technical support from L.~Huisman and O.~Böll.
\section{\label{sec:AuthorContributions}Author Contributions}
S.H., H.B., A.B., R.G.E.T., W.U., L.W. conceived the experiment. J.H. developed the laser system, performed the experiments, analyzed the data, and drafted the first version of the manuscript. J.H., A.T., R.B., T.F.. developed the vacuum system. I.T. and J.H. performed the trajectory simulations using an adapted version of a code originally written by A.T. and H.B. A.T. helped with an early version of the experiment and data analysis. S.H. and H.B. supervised the project. All authors discussed and approved the final manuscript. 

\section{\label{sec:CompetingInterests} Competing interests}
The authors declare no competing interests.

\bibliography{Papers_references}

\end{document}